\documentclass[aps,preprint,prb,showpacs,floatfix]{revtex4}

\usepackage{epsfig,amsmath,amssymb}
\usepackage{graphics} 
\usepackage{subfig}
\usepackage{color}

\usepackage{multirow}

\begin{document}

\title
{
Ground-state phases of the frustrated spin-$\frac{1}{2}$ $J_{1}$--$J_{2}$--$J_{3}$ Heisenberg 
ferromagnet ($J_{1}<0$) on the honeycomb lattice with $J_{3}=J_{2}>0$\\
}

\author
{P.~H.~Y.~Li and R.~F.~Bishop}
\affiliation
{School of Physics and Astronomy, Schuster Building, The University of Manchester, Manchester, M13 9PL, UK}

\author
{D.~J.~J.~Farnell}
\affiliation 
{Division of Mathematics, Faculty of Advanced Technology, 
University of Glamorgan, Pontypridd CF37 1DL, Wales, UK}

\author{J. Richter}

\affiliation{Institut f\"ur Theoretische Physik, Otto-von-Guericke 
Universit\"at Magdeburg, P.O.B. 4120, 39016 Magdeburg, Germany}

\author
{C.~E.~Campbell}
\affiliation
{School of Physics and Astronomy, University of Minnesota, 116 Church Street SE, Minneapolis, Minnesota 55455, USA}

\begin{abstract}

  We study the ground-state (gs) properties of the frustrated
spin-$\frac{1}{2}$ $J_{1}$--$J_{2}$--$J_{3}$ Heisenberg model on the two-dimensional 
honeycomb lattice with ferromagnetic nearest-neighbor ($J_{1}=-1$) 
exchange and frustrating antiferromagnetic next-nearest-neighbor ($J_{2}>0$)
and next-next-nearest-neighbor ($J_{3}>0$) exchanges, for the case $J_{3}=J_{2}$. 
We use the coupled-cluster method implemented to high orders of approximation, 
complemented by the Lanczos exact diagonalization of a large finite lattice with 32 sites,
in order to calculate the gs energy, magnetic order parameter,
and spin-spin correlation functions.  In one scenario we find a quantum
phase transition point between regions characterized by ferromagnetic
order and a form of antiferromagnetic (``striped'') collinear order 
at $J^{c}_{2} \approx 0.1095 \pm 0.0005$, which is below the corresponding hypothetical
transition point at $J^{\rm cl}_{2}=\frac {1}{7}$ ($\approx 0.143$) for the classical version of the 
model, in which we momentarily ignore the intervening noncollinear spiral 
phase in the region $\frac{1}{10}<J_{2}<\frac{1}{5}$.  
Hence we see that quantum fluctuations appear to stabilize somewhat the 
collinear antiferromagnetic order in preference to the ferromagnetic order in this model. 
We compare results for the present ferromagnetic case (with $J_{1}=-1$) 
to previous results for the corresponding antiferromagnetic case (with $J_{1}=+1$).  
The magnetic order parameter is found to behave similarly for the ferromagnetic 
and the antiferromagnetic models for large values of the frustration parameter 
$J_{2}$.  However, there are considerable differences in the behavior of the order 
parameters for the two models for $J_{2}/|J_{1}| \lesssim 0.6$.  For example,
the quasiclassical collinear magnetic long-range order for the antiferromagnetic
model (with $J_{1}=+1$) breaks down at $J^{c_{2}}_{2} \approx 0.60$,
whereas the ``equivalent'' point for the ferromagnetic model (with $J_{1}=-1$) 
occurs at $J^{c}_{2} \approx 0.11$. Unlike in the antiferromagnetic model (with $J_{1}=+1$),
where a plaquette valence-bond crystal phase intrudes between the two
corresponding quasiclassical antiferrmagnetic phases (with N\'eel and striped order)
for $J^{c_{1}}_{2} < J_{2} < J^{c_{2}}_{2}$, with $J^{c_{1}}_{2} \approx 0.47$,
we find no clear indications at all in the ferromagnetic model 
for an intermediate magnetically disordered phase between the corresponding 
phases exhibiting ferromagnetic and striped order.  Instead the evidence for the
ferromagnetic model (with $J_{1}=-1$) points to one of two scenarios:  either there 
is a direct first-order transition between the two magnetically ordered phases, as mentioned
above; or there exists an intervening phase between them in the very narrow range
$0.10 \lesssim J_{2} \lesssim 0.12$, which is probably a remnant of the spiral
phase that exists in the classical counterpart of the model over the larger range
$\frac{1}{10}<J_{2}<\frac{1}{5}$.

\end{abstract}


\maketitle

\section{Introduction}
\label{intro}

In recent years frustrated quantum spin systems on regular two-dimensional 
(2D) lattices  have aroused a great deal of research interest.
\cite{2D_magnetism_1,2D_magnetism_2,2D_magnetism_3}  In particular the 
interplay of magnetic frustration and quantum fluctuations has been 
seen to be a very effective route to destabilize or destroy
magnetic order and thereby to create new quantum phases.
Such 2D magnetic systems can thus in turn develop a diverse array of phases with 
widely different ordering properties, such as antiferromagnets with quasiclassical N\'eel
ordering, quantum ``spirals'', valence-bond
crystals/solids, phases with nematic ordering, and spin liquids.  Other factors that
influence the ground-state (gs) phase structures are the nature of the
underlying crystallographic lattice, the number and nature of the bonds on
this lattice, and the spin quantum numbers of the atoms localized to
the sites on the lattice.  The theoretical investigation of these
models has proceeded hand in hand with the discovery and experimental
investigation of ever more quasi-2D magnetic materials with novel properties.

One of the most intensively studied of all of the frustrated 2D models is the 
spin-$\frac{1}{2}$ $J_{1}$--$J_{2}$ Heisenberg antiferromagnet (HAF)  on the square lattice with
nearest-neighbor (NN) bonds (of strength $J_{1} > 0$) competing with
next-nearest-neighbor (NNN) bonds (of strength $J_{2} \equiv \alpha
J_{1} > 0$). This quantum system has two different quasiclassical phases with collinear
magnetic long-range order (LRO) at small ($\alpha < \alpha_{c_{1}}
\approx 0.4$) and large ($\alpha > \alpha_{c_{2}} \approx 0.6$) values
of the frustration strength parameter $\alpha$, 
separated by an intermediate quantum paramagnetic phase
with no magnetic LRO in the regime $\alpha_{c_{1}} < \alpha <
\alpha_{c_{2}}$.  Interest in this model has been greatly stimulated recently
by its experimental realization in such layered magnetic materials as
Li$_2$VOSiO$_4$,\cite{Mel:2000,Ro:2002}
Li$_2$VOGeO$_4$,\cite{Mel:2000} and VOMoO$_4$.\cite{Bom:2005}  The syntheses of such layered
quasi-2D materials has stirred up a great deal of renewed interest in
the model (and see also, e.g.,
Refs.~[\onlinecite{Ca:2000,Ca:2001,Sin:2003,Ro:2004}]).  Amongst
several methods that have been very successfully applied to 
the $J_{1}$--$J_{2}$ model has been the coupled cluster method 
(CCM),\cite{j1j2_square_ccm1,Bi:2008_PRB,Bi:2008_JPCM,j1j2_square_ccm5} which 
has also been applied to many similar strongly-interacting and 
highly frustrated spin-lattice models with comparable success. 
Other frustrated 2D models that have similarly engendered great recent 
interest include the spin-$\frac{1}{2}$ HAFs 
on the triangular\cite{Bernu:1994,Zeng:1995} and kagome
lattices.\cite{Kagome_Schn:2008,Eve:2010}

There has been a large amount of recent experimental investigation 
of the properties of quasi-2D magnetic materials with a
ferromagnetic (FM) NN coupling ($J_{1}<0$) and an antiferromagnetic
(AFM) NNN coupling ($J_{2} > 0$).  Examples include
Pb$_{2}$VO(PO$_{4}$)$_{2}$,\cite{Kaul:2004,Skoulatos:2007,Carretta:2009,Skoulatos:2009,Nath:2009}
(CuCl)LaNb$_{2}$O$_{7}$,\cite{Kageyama:2005}
SrZnVO(PO$_{4}$)$_{2}$,\cite{Skoulatos:2009,Tsirlin:2009_PRB79,Tsirlin:2009_PRB80}
BaCdVO(PO$_{4}$)$_{2}$,\cite{Carretta:2009,Tsirlin:2009_PRB79,Nath:2008}
PbZnVO(PO$_{4}$)$_{2}$,\cite{Tsirlin:2010} and
(CuBr)LaNb$_2$O$_7$.\cite{oba2006}  These experimental studies have also served to 
reignite interest in the theoretical investigation of the
gs and thermodynamic properties of the FM
$J_{1}$--$J_{2}$ model, i.e., the model with FM NN exchange ($J_{1}<0$)
and frustrating AFM NNN exchange ($J_{2}>0$).{\cite{shannon04,shannon06,sindz07,schmidt07,schmidt07_2,sousa,
shannon09,sindz09,haertel10,Richter:2010,sousa2011,momoi2011}
Interestingly, arguments for the existence of a spin-nematic phase between
two quasiclassical magnetically-ordered phases were
presented.\cite{shannon06,shannon09,sindz09,momoi2011} On the other hand, the
existence of such a non-classical magnetically-disordered phase was also 
questioned in Ref.~[\onlinecite{Richter:2010}].     

Other systems that have grown in importance in the last few years are
various spin-$\frac{1}{2}$ magnetic models defined on the 2D honeycomb lattice.
Several such systems have been both theoretically and experimentally studied
\cite{wang2010,honey1,honey2,honey3,honey5,schmidt2011,lauchli2011,honey7,oitmaa2011,exp}
intensively, partly because of their special properties and partly due to
the recent syntheses of various quasi-2D honeycomb-lattice materials.  
One reason for the theoretical interest in such models on the 2D honeycomb lattice is that a
spin-liquid phase has been found for the exactly solvable Kitaev
model,\cite{kitaev} in which the spin-$\frac{1}{2}$ particles reside
on just such a lattice.  Furthermore, the honeycomb lattice is obviously germane to
the very active research field of graphene, where the relevant physics
may well be described by Hubbard-like models on this
lattice.\cite{graphene,meng,schmidt2011} Interestingly, Meng {\it
  et al.}\cite{meng} found that for the Hubbard model on the honeycomb
lattice with moderate values of the Coulomb repulsion parameter $U$, strong
quantum fluctuations lead to an insulating spin-liquid phase
between the non-magnetic metallic phase and the AFM Mott insulator phase.
From the experimental side recent observations on the
spin-$\frac{3}{2}$ honeycomb-lattice HAF Bi$_3$Mn$_4$O$_{12}$(NO$_3$)
demonstrate a spin-liquid-like behavior at temperatures much lower
than the Curie-Weiss temperature.\cite{exp}

We have recently studied\cite{DJJF:2011_honeycomb} the AFM $J_{1}$--$J_{2}$--$J_{3}$ 
honeycomb model for the case where the spin quantum number $s$ of each 
of the spins on every lattice site is $s=\frac{1}{2}$, and with 
AFM nearest-neighbor exchange bonds ($J_{1}>0$) in the presence of
frustration caused by AFM NNN bonds ($J_{2}>0$) and with
next-next-nearest neighbor (NNNN) bonds of strength $J_{3}$ also present,
for the special case where $J_{3}=J_{2}$. We found\cite{DJJF:2011_honeycomb} 
that the scenario of deconfined criticality may hold for this model 
(and see also Ref.~[\onlinecite{lauchli2011}]).
To date there exist only limited studies of the corresponding
FM $J_{1}$--$J_{2}$--$J_{3}$ model (namely where $J_{1}<0$).  In this paper, we further the
investigation into the FM $J_{1}$--$J_{2}$--$J_{3}$
honeycomb model with FM NN bonds (of strength $J_{1}<0$) in the presence of frustrating
AFM NNN bonds (of strength $J_{2}>0$) and NNNN bonds (of strength $J_{3} > 0$).  
Once again we consider only the interesting special case where $J_{3}=J_{2}$.  
We focus our attention in the present study particularly on the 
detection and characterization of the gs phases of the
quantum model. Bearing in mind the controversial discussion of the
corresponding $J_{1}$--$J_{2}$ square-lattice  model with FM NN 
exchange bonds ($J_{1}<0$),  the question naturally arises  
as to whether any indications for  a non-classical
magnetically-disordered phase might now be found for the honeycomb model.
To determine the relevant gs phases and their properties we
calculate the gs energy, the spin-spin correlation function, and the magnetic order
parameter for the stripe-ordered state discussed below that is present 
as a gs phase in the corresponding 
classical version (equivalent to the $s \rightarrow \infty$ limit) of the model. 

In view of its proven past ability to give results of high accuracy for a wide 
variety of highly frustrated 2D spin-lattice models, we again use 
the coupled cluster method (CCM) as our main computational tool 
in this paper.  Additionally, we use the exact diagonalization (ED)
method for a large finite lattice of $N=32$ spins as a validity check of our CCM
results.  Since at the classical level the model now under consideration also exhibits
some similarities with the corresponding model with  AFM NN bonds ($J_{1}>0$), 
we compare our results for the quantum model of the FM case ($J_{1} \equiv -1$) 
with those of the corresponding AFM case ($J_{1} \equiv +1$).

The rest of the paper is organized as follows.  After describing the
model in Sec. \ref{model_section}, we apply the CCM to investigate its
gs properties. The CCM itself is very briefly described in
Sec. \ref{CCM}, before presenting and discussing our CCM and ED
results in Sec. \ref{results}.  We conclude in
Sec. \ref{conclusions} with a summary of the main results.

\section{The model}
\label{model_section}
The Hamiltonian of the spin-$\frac{1}{2}$ $J_{1}$--$J_{2}$--$J_{3}$
Heisenberg model on the honeycomb
lattice, which we studied
recently\cite{DJJF:2011_honeycomb} for the AFM case ($J_{1}>0$) is defined as
\begin{equation}
H = J_{1}\sum_{\langle i,j \rangle} \mathbf{s}_{i}\cdot\mathbf{s}_{j} + J_{2}\sum_{\langle\langle i,k \rangle\rangle} \mathbf{s}_{i}\cdot\mathbf{s}_{k} + 
J_{3}\sum_{\langle\langle\langle i,l \rangle\rangle\rangle} \mathbf{s}_{i}\cdot\mathbf{s}_{l}\,,   
\label{H}
\end{equation}
where $i$ runs over all lattice sites on the lattice, and where $j$ runs
over all NN sites connected to site $i$ by $J_{1}$ bonds, 
$k$ runs over all NNN sites connected to site $i$ by $J_{2}$ bonds, and 
$l$ runs over all NNNN sites connected to site $i$ by $J_{3}$ bonds, 
but counting each bond once and once only in the three sums. 
Each site $i$ of the lattice carries a spin-$\frac{1}{2}$
particle with spin operator ${\bf s}_{i} \equiv (s_{i}^{x},s_{i}^{y},s_{i}^{z})$.
We note that precisely the same model has also been studied recently on the square 
lattice, both in the case where all the bonds are AFM in nature,\cite{Reuther:2011_J1J2J3mod}
and in the FM case where $J_{1}<0$ and $J_{2}>0, J_{3}>0$.\cite{Feldner:2011_J1J2J3mod}

The aim of the present work is now to study further the 
spin-$\frac{1}{2}$ $J_{1}$--$J_{2}$--$J_{3}$ FM model (namely the above model in the case
$J_{1}<0$) on the honeycomb
lattice.\cite{honey1,honey2,honey3,honey4,honey5}  The lattice and
the exchange bonds are illustrated in Fig.~\ref{model}(a).
\begin{figure*}[!tb]
\mbox{
\subfloat[FM]{\scalebox{0.35}{\includegraphics{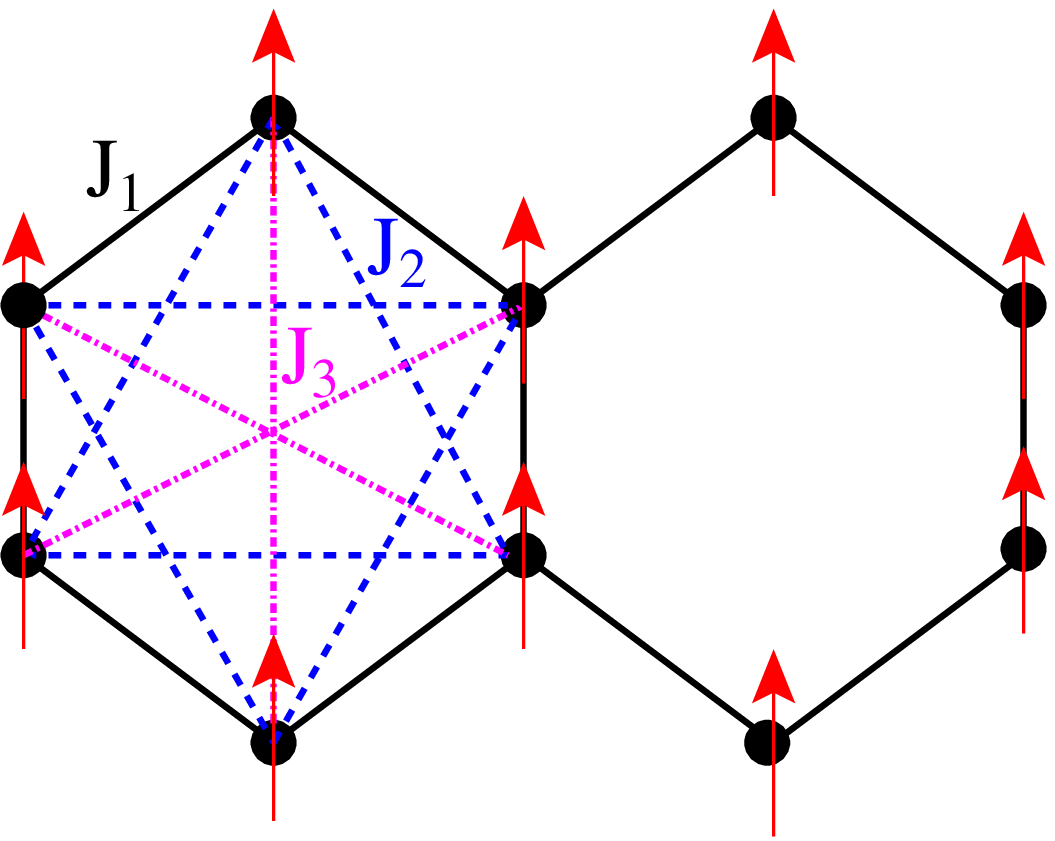}}}
\qquad
\subfloat[striped AFM]{\scalebox{0.35}{\includegraphics{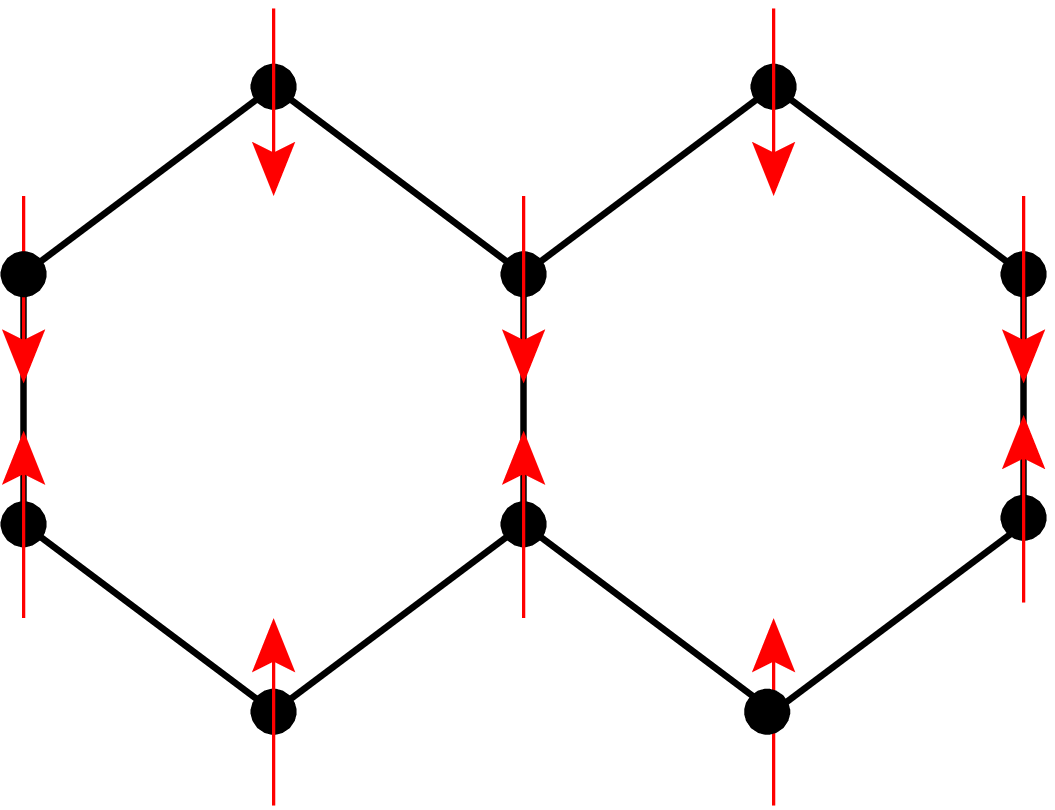}}}
}
\caption{(Color online) The $J_{1}$-$J_{2}$-$J_{3}$ honeycomb model, showing
  (a) the ferromagnetic (FM) state and (b) the classical striped antiferromagnetic (AFM) state. 
  The arrows represent spins located on lattice sites \textbullet.}
\label{model}
\end{figure*}
The classical gs phase diagram for the $J_{1}$--$J_{2}$--$J_{3}$
AFM model (with $J_{1}>0$) on the honeycomb lattice model displays 
collinear N\'{e}el and striped phases, both AFM in nature, as well as a spiral phase. These
phases meet in a triple point at $J_{3}=J_{2}=J_{1}/2$ (and for more
details see, e.g., Ref.~[\onlinecite{honey3}]).  For the
remainder of this paper we again focus on the case where $J_{3}=J_{2}>0$}, but now where the 
NN exchange bond is FM in nature ($J_{1}<0$).  The
gs energies of the only two corresponding classical collinear states are
then given by
\begin{eqnarray}
\frac{E^{\mbox{\tiny{cl}}}_{\mbox{\tiny{FM}}}}{{N}} &=& s^{2} \left (\frac{3}{2}J_{1} + \frac{9}{2}J_{2} \right),\nonumber \\ 
\frac{E^{\mbox{\tiny{cl}}}_{\mbox{\tiny{striped AFM}}}}{{N}} &=& s^{2} \left (\frac{1}{2}J_{1} - \frac{5}{2}J_{2} \right), \label{H_classical}
 \end{eqnarray}
for the FM state and collinear striped AFM state shown in Figs.~\ref{model}(a) and (b) respectively. 
If these were the only gs phases in this $J_{1}<0$ regime we would thus 
have a classical transition between the FM state and the striped AFM state at $J^{\rm cl}_{2}=-\frac{1}{7} J_{1}$
($\approx 0.143$ for $J_{1}=-1$) and a classical energy per site at this point of
$E^{\rm cl}/N = -\frac{3}{14} \approx -0.214$ for the $s=\frac{1}{2}$ system with $J_{1}=-1$.
For the corresponding AFM model with $J_{1}>0$ such a  striped AFM
state also exists as stated above, but the classical transition
between the AFM N\'{e}el state and the striped AFM state is at
$J_{2}=+\frac{1}{2}J_{1}$.  The reason why the corresponding phase
transition in the FM model occurs at a smaller value of the frustration
parameter, $J^{\rm cl}_{2}/|J_{1}|=\frac{1}{7}$ than the value
$J^{\rm cl}_{2}/J_{1}=\frac{1}{2}$ for the AFM model is due
to the $J_{3}$($>0$) NNNN exchange bonds that act to frustrate the 
fully polarized FM state, whereas they reinforce the AFM N\'eel state.
By contrast, the $J_{2}$($>0$) NNN exchange bonds act to frustrate
the $J_1$ bonds for both the FM state of the FM model and the
N\'eel state of the AFM model.
We note that the classical FM state is also an eigenstate of the Hamiltonian, with 
energy eigenvalue equal to the energy of its classical FM counterpart.

We note, however, that in fact the classical $J_{1}$--$J_{2}$--$J_{3}$ Heisenberg model on 
the honeycomb lattice with $J_{1}<0$, $J_{2}>0$, and $J_{3}>0$ also has a spiral
phase that intervenes in a very narrow strip between the FM phase and the collinear 
striped AFM phase.  (In Refs.~[\onlinecite{honey3,Rastelli:1979}] this is referred to
as phase V.)  The region in the $x$-$y$ plane (where $x \equiv J_{2}/J_{1}$ and $y \equiv J_{3}/J_{1}$) 
in which it is the stable gs phase in the case $J_{1}<0$ is bounded by the three
curves (i) $y=0$, $-\frac{1}{2}<x<-\frac{1}{6}$, (ii) $y=-\frac{3}{2}x-\frac{1}{4}$,
$-\frac{1}{6}<x<\frac{1}{2}$, and (iii) $y=\frac{1}{8}[1-6x-\sqrt{36x^2+20x+17}]$,
$-\frac{1}{2}<x<\frac{1}{2}$.  The point $(x=\frac{1}{2},y=-1)$ is a classical
tetracritical point at which the spiral phase V meets the FM phase, the striped
collinear AFM phase, and the AFM N\'eel phase (and see Fig. 3 of 
Ref.~[\onlinecite{honey3}] for further details).  Thus, in our case, where
$J_{3}=J_{2}$ and $J_{1}<0$, the classical spiral phase V exists in the
narrow region $\frac{1}{10}<J_{2}/|J_{1}|<\frac{1}{5}$.  Naturally this 
includes the point $J^{\rm cl}_{2}/|J_{1}|=\frac{1}{7}$ discussed above
at which the FM and striped collinear AFM phases would meet in the
absence of the spiral phase V as a stable gs phase intervening between them.

In all of our results below for the FM $J_{1}$--$J_{2}$--$J_{3}$ 
honeycomb 
system we henceforth set $J_{1} \equiv -1$
with no loss of generality, since this simply sets the overall scale of the
Hamiltonian, and we consider the case where $J_{3}=J_{2}>0$, such
that both the NNN and NNNN bonds act to frustrate the ferromagnetism.

\section{Coupled Cluster Method}
\label{CCM}
The CCM (see, e.g., Refs.~[\onlinecite{Bi:1991,Ze:1998,Fa:2004}] and
references cited therein) that we use here is one of the most powerful
and most versatile modern techniques in quantum many-body theory.  It
has been used to study various quantum magnets (see, e.g., 
Refs.~[\onlinecite{j1j2_square_ccm1,Bi:2008_PRB,Bi:2008_JPCM,j1j2_square_ccm5,Fa:2001,Kr:2000,
Schm:2006,Fa:2004,Ze:1998,shastry2,shastry3,square_triangle,UJack_ccm,Bi:2010_kagomeSquare,Bi:2011_UJack_greatSpins}]
and references cited therein) very successfully.  The method is
particularly suitable for investigating frustrated systems, due to the
fact that some of the main alternative methods are restricted by
certain problems that arise in such cases.  For instance, quantum Monte Carlo (QMC) techniques
suffer from the infamous and well-known ``sign problem'' for such systems.  The exact
ED method is also usually restricted by available computational
power to relatively small finite-sized lattices.  Nevertheless it
can often be used, as here, to provide a handy tool to
check and validate the results of other numerical or approximate
methods.

We briefly describe here some of the important features of the CCM 
as applied to spin-lattice problems (and see, e.g., 
Refs.~[\onlinecite{Ze:1998,Kr:2000,Fa:2001,Schm:2006,j1j2_square_ccm1,Fa:2004}]
and references cited therein for further details).  The starting point for any
CCM calculation is to select a normalized state
$|\Phi\rangle$ as a reference or model state against which to incorporate in a 
systematic and potentially exact fashion the correlations present in the 
exact ground state.  We often use a relevant classical ground state as the model 
state for spin systems for the sake of convenience, but other appropriate states
may certainly also be used.  In order to treat each site
equivalently, a mathematical rotation of the local axes of
the spins is conveniently performed in such a way that all spins in the 
reference state align along the same direction, say the
negative $z$-axis.  Clearly, such rotations leave unchanged the SU(2) 
commutation relations between components of the spin operators.  

The exact ket and bra gs energy eigenstates, 
$|\Psi\rangle$ and $\langle\tilde{\Psi}|$, of the many-body system are
then parametrized in the CCM form as: \vskip0.2cm
\begin{equation}
|\Psi\rangle = \mbox{e}^{S}|\Phi\rangle; \qquad S = \sum_{I\neq0}{\cal S}_{I}C^{+}_{I},   \label{eq:ket_eq}
\end{equation}
\begin{equation}
\langle\tilde{\Psi}| = \langle\Phi|\tilde{S}\mbox{e}^{-S}; \qquad \tilde{S} = 1 + \sum_{I\neq0}\tilde{{\cal S}_{I}}C^{-}_{I},   \label{eq:bra_eq}
\end{equation}
where
\begin{equation}
H|\Psi\rangle = E|\Psi\rangle; \qquad \langle\tilde{\Psi}|H = E\langle\tilde{\Psi}|,  \label{eq:SE_CCM}
\end{equation}
are the Schr\"{o}dinger gs ket and bra equations respectively.
The multiconfigurational creation operators $C^{+}_{I} \equiv (C^{-}_{I})^{\dagger}$ are
defined so that $\langle \Phi |C^{+}_{I} = 0 = C^{-}_{I}| \Phi \rangle \;; \forall I \neq 0$, and
where we have defined $C^{+}_{0} \equiv 1 \equiv C^{-}_{0}$. 
They are required to form a complete set of mutually commuting many-body creation operators 
in the Hilbert space, defined with respect to $|\Phi\rangle$ as a cyclic vector. 
Clearly the states are normalized such that $\langle\tilde{\Psi}|\Psi\rangle =
\langle \Phi| \Psi\rangle = \langle \Phi| \Phi \rangle \equiv 1$.
For spin-lattice systems they take the form of multi-spin raising operators and 
are written as products of single-spin raising operators, $C^{+}_{I}\equiv s^{+}_{j_{1}}
s^{+}_{j_{2}} \cdots s^{+}_{j_{n}}$, where $s^{+}_{j} \equiv s^{x}_{j} + is^{y}_{j}$.  
The gs energy is calculated in terms of the correlation 
coefficients $\{{\cal S}_{I}\}$ as $E=\langle\tilde{\Psi}|H|\Psi\rangle = 
\langle\Phi|\mbox{e}^{-S}H\mbox{e}^{S}|\Phi\rangle$; and the average on-site
magnetization $M$ in the rotated spin coordinates is calculated 
equivalently in terms of the coefficients $\{{\cal S}_{I},\tilde{{\cal S}_{I}}\}$ 
as $M \equiv -\frac{1}{N} \langle\tilde{\Psi}|\sum_{j=1}^{N}s^{z}_{j}|\Psi\rangle$, 
which now plays the role of the order parameter.
Finally, the complete set of unknown ket- and bra-state correlation coefficients 
$\{{\cal S}_{I}, \tilde{{\cal S}_{I}}\}$ is calculated by requiring the expectation
value $\bar{H}=\langle\tilde{\Psi}|H|\Psi\rangle$ to be a minimum with
respect to all parameters $\{{\cal S}_{I}, \tilde{{\cal S}_{I}}\;; \forall I \neq 0\}$.
This readily leads to the coupled set of nonlinear equations for the ket-state creation 
correlation operators $\{{\cal S}_{I}\}$, $\langle \Phi|C^{-}_{I}\mbox{e}^{-S}H\mbox{e}^{S}|\Phi\rangle =
0\;; \forall I \neq 0$, and to the coupled set of linear equations, 
$\langle\Phi|\tilde{S}(\mbox{e}^{-S}H\mbox{e}^{S} - E)C^{+}_{I}|\Phi\rangle = 0\;; \forall I \neq 0$, 
which can then be solved for the bra-state destruction correlation operators
$\{\tilde{{\cal S}_{I}}\}$.  

When all many-body configurations $\{I\}$ are included in the expansions of the correlation expansions
operators $S$ and $\tilde{S}$, the CCM formalism is exact.  However, it is
necessary of course in practice to use approximation schemes to truncate the 
sets of configurations $\{I\}$ contained in the expansions of 
Eqs. (\ref{eq:ket_eq}) and (\ref{eq:bra_eq}) for the CCM  
correlation operators.  For systems defined on a regular periodic
spatial lattice as here, it is convenient to use the well-established
LSUB$m$ approximation scheme in which all possible multi-spin-flip
correlations over different locales on the (here, honeycomb) lattice defined by
$m$ or few contiguous lattice sites are retained.  
Clusters are defined to be contiguous in this sense if every 
site in the cluster is adjacent (as a nearest neighbor) 
to at least one other site in the cluster.  This is the scheme 
we use for all our results presented below.  The number $N_{f}$ of
independent fundamental clusters (i.e., those that are inequivalent
under the symmetries of the Hamiltonian and of the model state)
increases rapidly with the truncation index $m$, as shown in 
Table~\ref{table_FundConfig}
\begin{table}
  \caption{Number of fundamental LSUB$m$ configurations ($N_{f}$) for the collinear 
   striped AFM state of the spin-$\frac{1}{2}$ $J_{1}$--$J_{2}$--$J_{3}$ honeycomb model.}
\label{table_FundConfig}
\begin{tabular}{cc} \hline\hline
{Method} & {$N_{f}$} \\ 
\hline 
LSUB6 & 72    \\ 
LSUB8 & 941   \\ 
LSUB10 & 14679 \\
LSUB12 & 250891 \\ \hline\hline
\end{tabular} 
\end{table}
for the present spin-$\frac{1}{2}$ $J_{1}$--$J_{2}$--$J_{3}$
model on the honeycomb lattice, where we use the natural lattice geometry
itself to define the notion of adjacency inherent in the definition 
of the LSUB$m$ scheme.  We see, for example, that the 
number $N_{f}$ of such fundamental clusters (and hence the number of simultaneous nonlinear
equations we need to solve for the retained correlation coefficients 
$\{{\cal S}_{I}\}$) for the striped model state is 250891 at the highest LSUB12 level of 
approximation that we utilize here.  The corresponding numbers, $N_f$, of
fundamental configurations are appreciably higher at a given LSUB$m$ level
when the spiral phase V is used as the CCM model state, due to the considerably
reduced symmetry.  It is necessary to use
massive parallelization and supercomputing resources in order to
perform the CCM calculations at such high level of approximation.\cite{ccm}  Thus, for 
example, to obtain a single data point (i.e., for a given value of $J_{3}=J_{2}$) for
the striped model state at the LSUB12 level typically requires about 0.5 h
computing time using 1000 processors simultaneously.  

We present CCM results below based on the striped collinear AFM state as model state,
at various LSUB$m$ levels of
approximation with $m=\{6,8,10,12\}$, and also in the corresponding $m \rightarrow \infty$
extrapolated limits (LSUB$\infty$) based on the
well-tested extrapolation schemes described below and in more detail
elsewhere.\cite{Ze:1998,Fa:2004,j1j2_square_ccm1,Bi:2008_PRB,Bi:2008_JPCM}
We have also performed extrapolations for the data set with
$m=\{6,8,10\}$.  Both sets of results agree well with 
one another, which gives added credence to our results.  
Note that we do not use the LSUB$m$ approximation scheme for values $m<6$ of
the truncation index, since these low-order approximations
will not capture the natural hexagonal structure of the lattice.
We remark that, as always, the CCM exactly obeys the
Goldstone linked-cluster theorem at every LSUB$m$ level of
approximation.  Hence we work from the outset in the limit $N
\rightarrow \infty$, where $N$ is the number of sites on the honeycomb 
lattice, and extensive quantities such as the gs energy are always
guaranteed to be linearly proportional to $N$ in this limit.
                                                                      
We clearly do not need to perform any finite-size scaling of our results, as all CCM
approximations are automatically performed from the outset in the
infinite-lattice limit, $N \rightarrow \infty$, as discussed above.  
It is, however, necessary to extrapolate to
the exact $m \rightarrow \infty$ limit in the LSUB$m$ truncation index
$m$, in which limit the complete (infinite) Hilbert space is reached.  For
the gs energy per spin, $E/N$, a well-tested and very accurate extrapolation
ansatz (and see, e.g., Refs.,
[\onlinecite{Bi:2008_PRB,Bi:2008_JPCM,Richter:2010,Kr:2000,rachid05,ccm3}])
is
\begin{equation}
E(m)/N = a_{0}+a_{1}m^{-2}+a_{2}m^{-4}\,,   \label{E_extrapo}
\end{equation}
while for the magnetic order parameter, $M$, different schemes have been
employed for different situations. For models showing no or only
relatively small amounts of frustration, a well-tested and accurate
rule (and see, e.g., Refs.~[\onlinecite{Kr:2000,rachid05}]) is
\begin{equation}
M(m) = b_{0}+b_{1}m^{-1}+b_{2}m^{-2}\,.    \label{M_extrapo_standard}
\end{equation}
For highly frustrated systems, particularly those showing a gs order-disorder transition, a more appropriate
extrapolation rule with fixed exponents that has been found to give good results (and see, e.g.,
Refs.~[\onlinecite{Bi:2008_PRB,Richter:2010}]) is
\begin{equation}
M(m) = c_{0}+c_{1}m^{-1/2}+b_{2}m^{-3/2}\,.    \label{M_extrapo_frustrated}
\end{equation}
We give illustrations here of the use of each of these schemes, wherever and whenever possible.

\section{Results and Discussion}
\label{results}
\begin{figure*}[!t]
\begin{center}
\mbox{
  \subfloat[$m=6,8,10,12$]{\scalebox{0.3}{\includegraphics[angle=270]{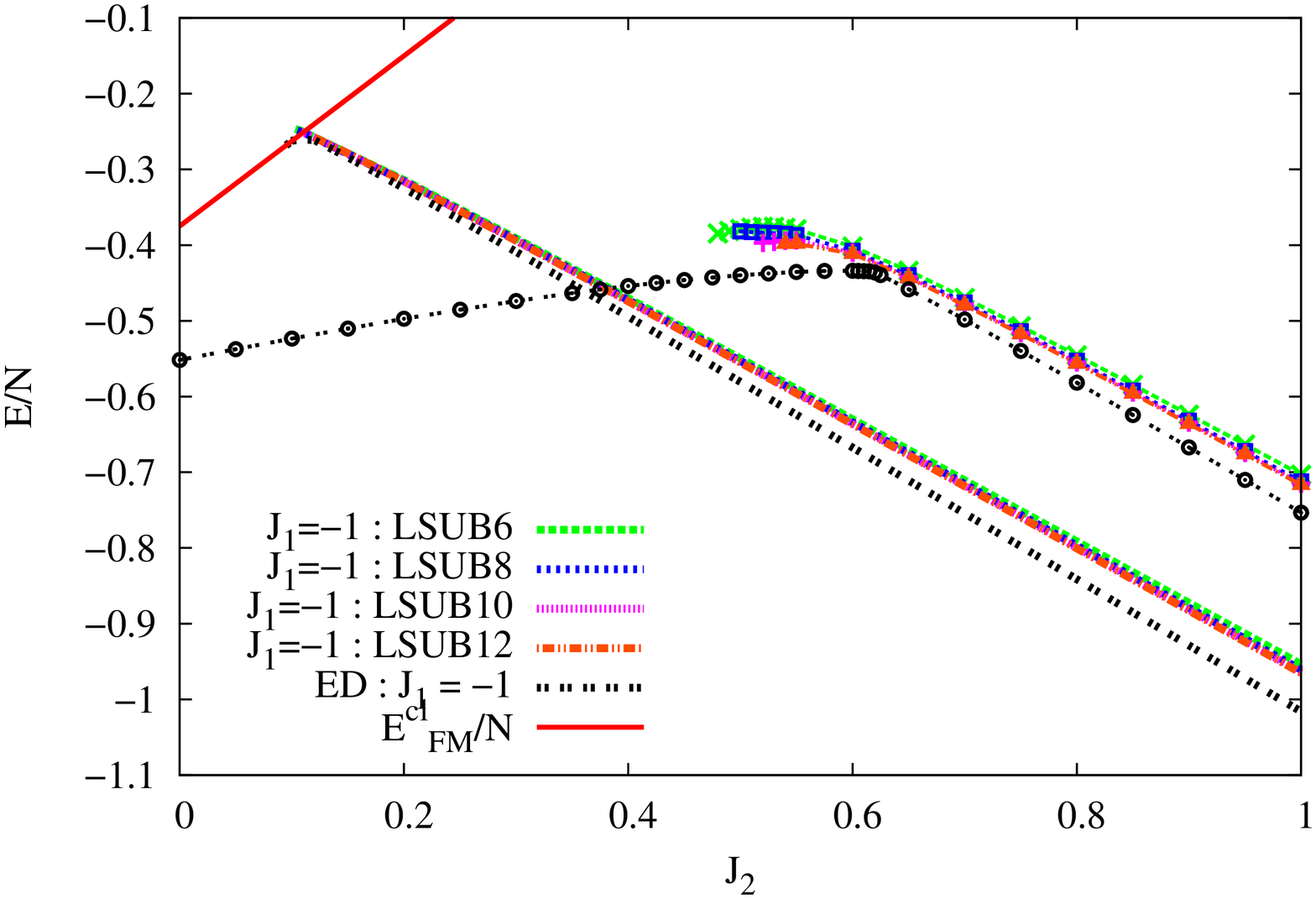}}}
 }
\mbox{
\subfloat[$m \rightarrow \infty$]{\scalebox{0.3}{\includegraphics[angle=270]{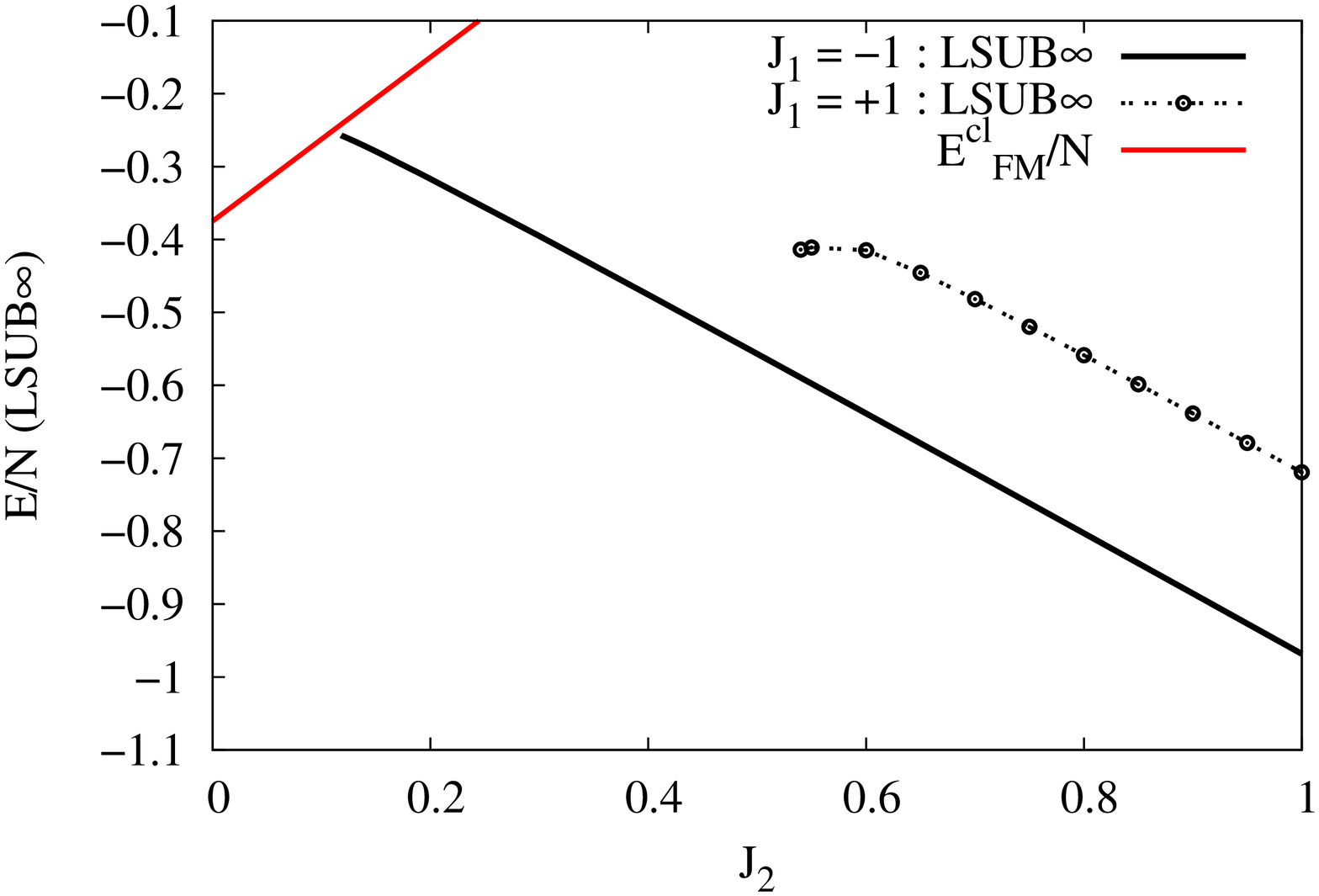}}}
}
\caption{
  (Color online) Ground-state energy per spin, $E/N$, versus
  $J_{2}$ for the striped phase of the spin-$\frac{1}{2}$
  $J_{1}$--$J_{2}$--$J_{3}$ honeycomb model with $J_{1}=-1$ compared with those for
  $J_{1}=+1$ (with $J_{3}=J_{2}>0$ for both cases).  The CCM results using
  the striped model state for various LSUB$m$ approximations with
  $m=\{6,8,10,12\}$, together with the ED ($N=32$) results are shown in (a).  
  We show in (b) the extrapolated CCM LSUB$\infty$ results using the $m=\{6,8,10,12\}$
  data points fitted to Eq. (\ref{E_extrapo}).  In all cases curves without symbols attached 
  refer to the case $J_{1}=-1$, whereas the corresponding curves 
  with symbols refer to the case $J_{1}=+1$. The FM result
  from Eq. (\ref{H_classical}) with $s=\frac{1}{2}$ is also displayed.
  }
\label{E}
\end{center}
\end{figure*}

We now present and discuss our CCM results.  In order to have an
independent check on the accuracy and consistency of our CCM results, 
we have also performed additional computations of selected gs properties of the
present models using the ED technique that is a well-established and
successful tool for studying frustrated quantum spin systems (and see, e.g.,
Refs.~[\onlinecite{Richter:2010,schulz,
waldtmann,ED40,lauchli2011,nakano2011}]).
In Fig.~\ref{E}(a) we show the CCM results for the gs energy per spin, $E/N$, in
various LSUB$m$ approximations based on the striped state as CCM model state, 
as well as the exact gs energy for a finite lattice of size $N=32$.
We also show separately, in Fig.~\ref{E}(b), the extrapolated ($m \rightarrow \infty$) results
obtained from Eq.~(\ref{E_extrapo}) using the data set
$m=\{6,8,10,12\}$.  Comparison is made with the results for the corresponding
AFM version of the model with $J_{1}=+1$.

The CCM LSUB$m$ data displayed in Fig.~\ref{E}(a) show that the gs energy results 
converge extremely rapidly as the truncation index $m$ is increased, such that 
the difference between the LSUB12 results and the extrapolated 
($m \rightarrow \infty$) results obtained from Eq. (\ref{E_extrapo}) is 
very small indeed.  We note too that, just as in the corresponding AFM case of the 
model with $J_{1}=+1$, the various CCM LSUB$m$ solutions based on the striped model 
state now also terminate at some lower termination point $J^{t}_{2}$ as $J_{2}$ is decreased.
Such terminations of CCM solutions are very common
and have been very well documented.\cite{Fa:2004} In all such cases a termination
point always arises due to the solution of the CCM
equations becoming complex at this point, beyond which
there exist two branches of entirely unphysical complex conjugate
solutions.\cite{Fa:2004}  In the region where the solution reflecting
the true physical solution is real there actually also exists
another (unstable) real solution. However, only the shown
branch of these two solutions reflects the true (stable) physical
ground state, whereas the other branch does not. The
physical branch is usually easily identified in practice as the
one which becomes exact in some known (e.g., perturbative)
limit. This physical branch then meets the corresponding unphysical
branch at some termination point with infinite
slope, beyond which no real solutions exist. The
LSUB$m$ termination points are themselves also reflections of
the quantum phase transitions in the real system and may be
used to estimate the position of the phase boundary,\cite{Fa:2004} although
we do not do so for this critical point in the FM model, since we
have more accurate criteria that we now discuss.

We note first from Fig.~\ref{E}(a) that the LSUB$m$ termination points using 
the striped state as the CCM model state for the
present FM version of the model with $J_{1}=-1$, lie very close indeed to 
the points where the curves cross (or nearly cross) the corresponding curve
for the FM state given by Eq. (\ref{H_classical}).  This gives us our
first evidence that either there is no intermediate phase between the quantum striped 
phase and the FM phase for the case $J_{1}=-1$, or, if one exists, it can occur 
only over a very narrow regime indeed.  This situation may be contrasted
with that of the AFM version of the model ($J_{1}=+1$),\cite{DJJF:2011_honeycomb} 
where the LSUB$m$ results for the
gs energy using the striped model state terminate before they meet the
corresponding results using the N\'eel state as model state (which themselves 
also terminated at some upper termination points that were lower in value than the 
lower termination points for the striped state).  In the latter case there is
an intermediate plaquette valence-bond crystal (PVBC) phase.  

At the classical level
the difference in the values of the gs energy per spin of the collinear striped states 
between the two $s=\frac{1}{2}$ cases (i.e., for positive and negative values of $J_{1}$ with
$|J_{1}|=1$) is 0.25, independent of $J_{2}$ and $J_{3}$.  The quantum versions
follow this pattern for larger values of $J_{3}=J_{2}$, as seen from  Fig.~\ref{E}, 
but the constancy in the difference breaks down at around $J_{2} \approx 0.6$, where the AFM case 
($J_{1}=+1$) exhibits a critical point marking a transition to the PVBC phase, which then in turn
undergoes a further phase transition to the N\'eel phase at another lower critical value.  The
corresponding best available CCM estimates for those two critical values for the AFM 
case of the model with $J_1=+1$ are $J^{c_{2}}_{2} \approx 0.60$ and $J^{c_{1}}_{2} \approx 0.47$ 
respectively.\cite{DJJF:2011_honeycomb}   In the present FM case of the model with
$J_1=-1$ we see no evidence (apart from the seeming termination of the solutions
to the equations for the LSUB$12$ approximation based on the striped state as
CCM model state very slightly before the gs energy crossing point with the FM state) of any
similar intermediate state between the FM state and the collinear striped AFM state.  If any
such intermediate state exists at all, however, it must be confined to a very narrow region indeed
around $J_{2} \approx 0.11$, probably confined to $0.10 \lesssim J_{2} \lesssim 0.12$.  We
return to a more detailed discussion of this region later.  For the moment we note only that
it is much reduced from the region $0.1<J_{2}<0.2$ in which the corresponding classical
version of the model has the spiral phase V as its stable gs phase.

For the present FM case with $J_{1}=-1$ the CCM LSUB$m$ gs energy curves using the striped model
state cross the corresponding gs energy curve for the FM state from Eq. (\ref{H_classical}) for
$m=\{6,8,10\}$ at corresponding critical values $J^{c}_{2}(\rm{LSUB}6) \approx 0.1106$ 
(where $E^{c}_{\rm{LSUB}6}/N \approx -0.2506$), $J^{c}_{2}(\rm{LSUB}8) \approx 0.1101$ 
(where $E^{c}_{\rm{LSUB}8}/N \approx -0.2511$), and $J^{c}_{2}(\rm{LSUB}10) \approx 0.1098$ 
(where $E^{c}_{\rm{LSUB}10}/N \approx -0.2515$).  The corresponding LSUB$12$ result for the gs energy
using the striped state as the CCM model state appears to terminate just before meeting the
gs energy curve for the FM phase.  However we note that for such very high-order CCM 
calculations it becomes very computationally expensive to determine the termination point with 
high accuracy.  If we use the extrapolated LSUB$\infty$ results for the gs energy for the striped
phase by making use of Eq. (\ref{E_extrapo}) and employing the whole data set $m=\{6,8,10,12\}$,
we thus need to perform a further very small extrapolation of the CCM results to lower values
of $J_{2}$ to find the presumed crossing point of the energies of the striped and FM phases, in
the scenario in which these two phases meet at a first-order transition with no intermediate 
phase (that would itself be confined to the very narrow intervening region 
$0.10 \lesssim J_{2} \lesssim 0.12$, as discussed above).  
As expected, simple power-law expansions give very accurate fits, and give crossing points
very close to those above.  Putting all the energy data together, our best estimate for the
critical point of the first-order phase transition from the collinear striped phase to the
FM phase (in the scenario where this transition occurs directly, with no intermediate phase 
confined to the narrow region $0.10 \lesssim J_{2} \lesssim 0.12$) 
for the spin-$\frac{1}{2}$ $J_{1}$--$J_{2}$--$J_{3}$ Heisenberg 
ferromagnet (with $J_{1}=-1$) on the honeycomb lattice, and with $J_{3}=J_{2}>0$, is
$J^{c}_{2}=0.1095 \pm 0.0005$, at which point the gs energy per spin is 
$E^{c}/N=-0.2518 \pm 0.0006$.

We see from Fig.~\ref{E}(a) that
the agreement between the ED ($N=32$) and the CCM energies is very satisfactory.  
Moreover, due to the finite-size scaling of the gs energy, $E/N = e_0-a/N^{3/2}$
with $a >0$ (and see, e.g. Refs.~[\onlinecite{schulz}] and [\onlinecite{ED40}]), 
the difference between the CCM and the ED gs energies would become even smaller 
if finite lattices of larger size could be considered. The ED turnover point in 
the energy curve that marks the termination of the FM phase occurs at a value of about 
0.1003 for the $N=32$ lattice used, and for the same reasons as above
this value will increase as $N$ is increased.  Thus, in summary, while the
CCM estimates for the gs energy per spin for the spin-$\frac{1}{2}$ $J_{1}$--$J_{2}$--$J_{3}$ 
Heisenberg model on the extended infinite honeycomb lattice 
are much more accurate than the ED results, the latter do serve as an
independent check on the former.

The hypothetical phase transition (i.e., when the existence of the intervening spiral phase V
is momentarily ignored) from FM order to collinear striped AFM order for the classical 
version of the FM model with $J_{1}=-1$ occurs at a value
$J^{\rm cl}_{2}=\frac{1}{7} \approx 0.143$, compared with the corresponding
value $J^{c}_{2} \approx 0.110$ found here.  Thus quantum fluctuations act to 
stabilize the collinear AFM order, at the expense of the FM order, to higher values 
of frustration than in the classical case.  It is interesting to note that 
a similar situation was found in the FM version 
($J_{1}=-1, J_{2}>0$) of the spin-$\frac{1}{2}$ $J_{1}$--$J_{2}$ 
model on the square lattice,\cite{Richter:2010} where a quantum critical point exists
at a value $J^{c}_{2} \approx 0.39$  for a similar transition from FM order to
collinear striped order, compared with a corresponding classical value of 
$J^{\rm cl}_{2}=0.5$.  It is well known, from many cases studied, that quantum
fluctuations almost always favor collinear states over noncollinear ones (e.g., 
spiral or canted states).  What is interesting in both the present case and the 
spin-$\frac{1}{2}$ $J_{1}$--$J_{2}$ model on the square lattice cited above, is that
quantum fluctuations seem also to favor one collinear state (namely the collinear striped
AFM state in these two cases) where the quantum fluctuations are present, 
over another collinear state (namely the FM state in these two cases) where quantum 
fluctuations are absent.  It is intriguing to wonder whether these are examples of 
a more general rule.

\begin{figure*}[!t,b]
\begin{center}
\mbox{
  \subfloat[$m=\{6,8,10,12\}$]{\scalebox{0.3}{\includegraphics[angle=270]{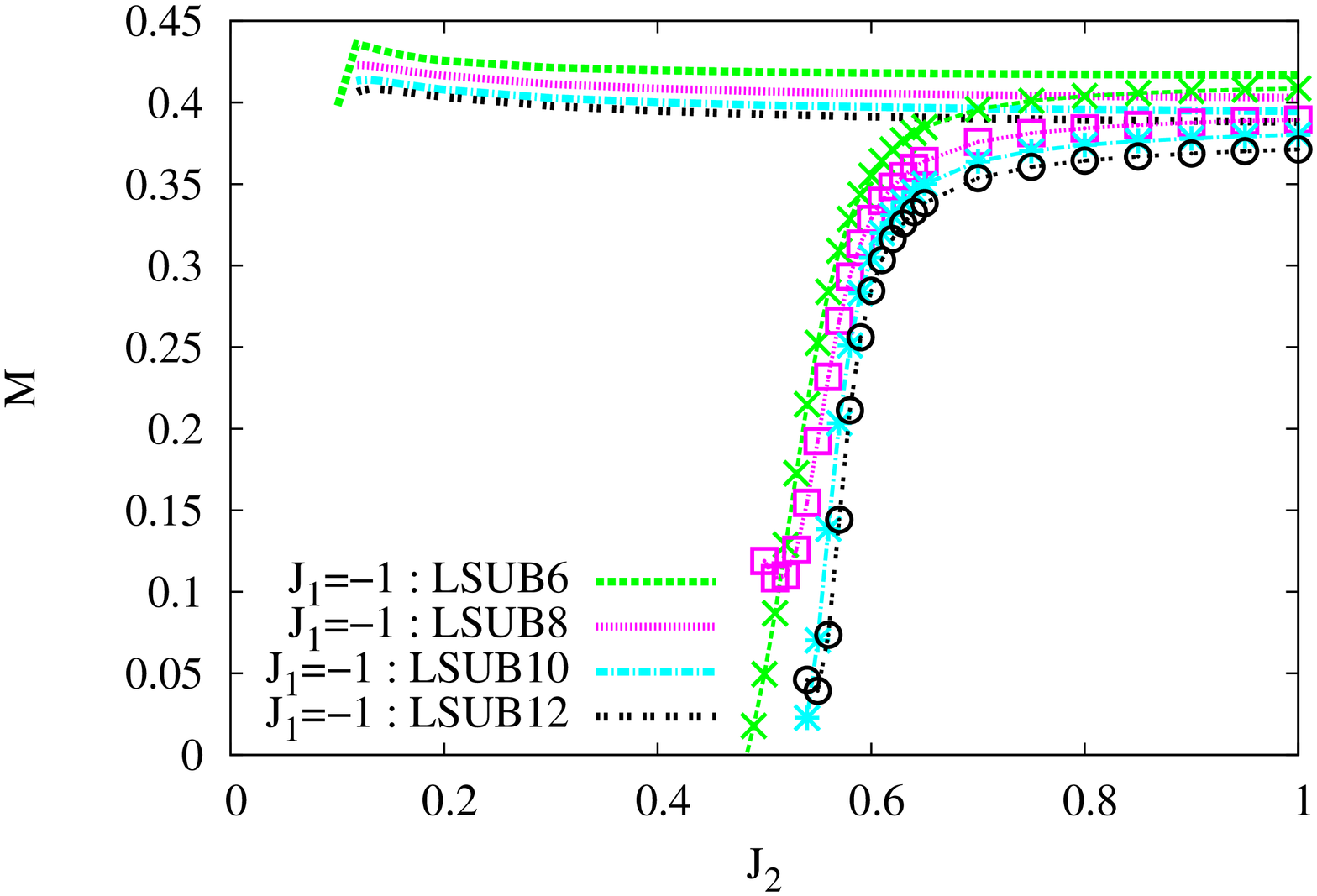}}}
\hfil
\subfloat[$m \rightarrow \infty$]{\scalebox{0.3}{\includegraphics[angle=270]{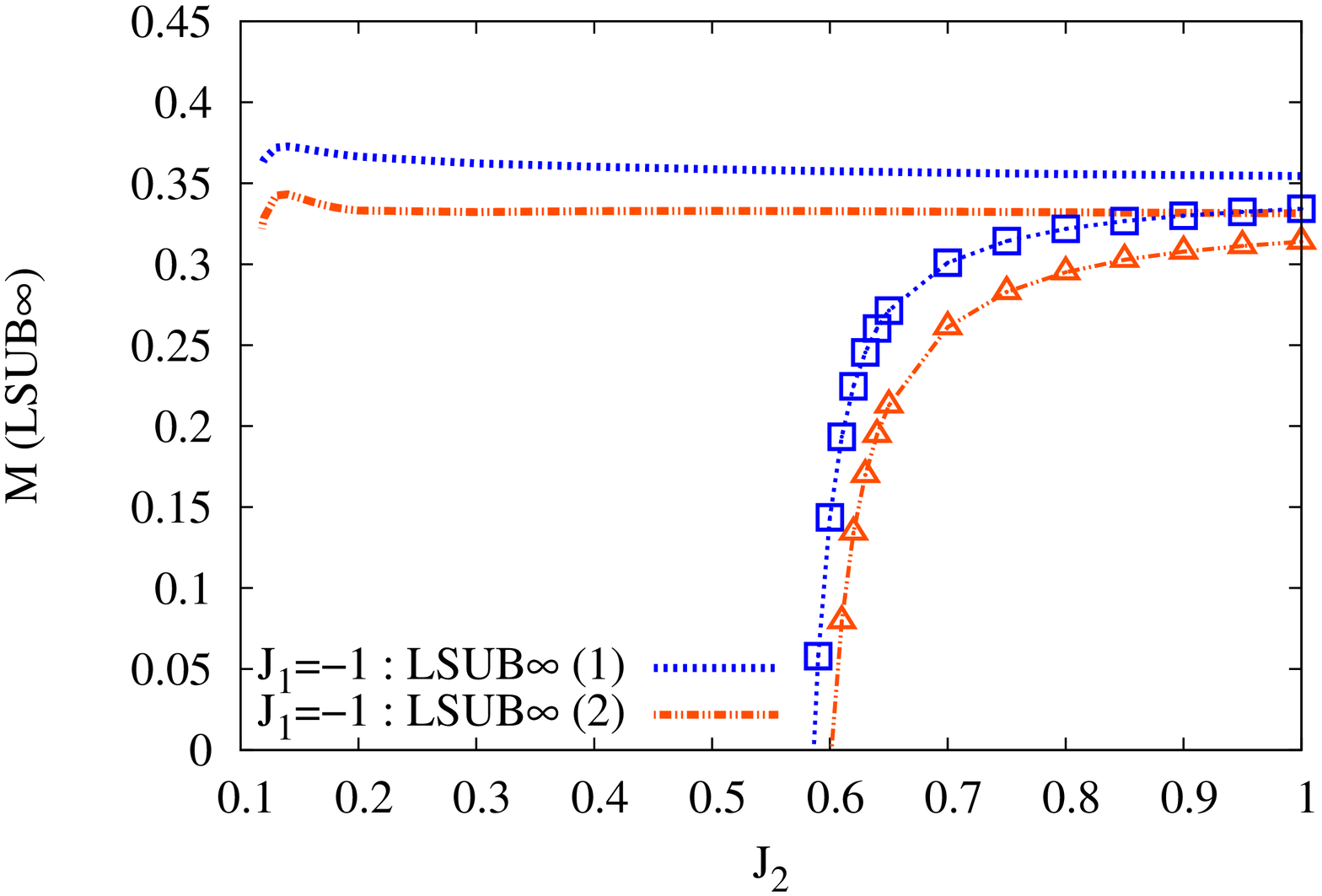}}}
 }
 \caption{(Color online) Ground-state magnetic order parameter, $M$, 
   for the striped AFM state of the spin-$\frac{1}{2}$ $J_{1}$--$J_{2}$--$J_{3}$
   honeycomb model with $J_{1}=-1$, compared with those for $J_{1}=+1$ (with $J_{3}=J_{2}>0$
   for both cases). The CCM results for
   various LSUB$m$ approximations with $m=\{6,8,10,12\}$ are shown in
   (a).  We also show in (b) the extrapolated LSUB$\infty$ results 
   using the $m=\{6,8,10,12\}$ data points.  The curves labelled LSUB$\infty (1)$ and LSUB$\infty (2)$
   are obtained using Eq. (\ref{M_extrapo_standard}) and Eq. (\ref{M_extrapo_frustrated})
   respectively.  In all cases curves without symbols attached refer to the case $J_{1}=-1$,
   whereas the corresponding curves with symbols refer to the case $J_{1}=+1$.}
\label{M}
\end{center}
\end{figure*}

We present results in Fig.~\ref{M} for the CCM collinear stripe order parameter $M$, as 
defined in Sec.~\ref{CCM}.  Figure~\ref{M}(a) shows LSUB$m$ results with $m=\{6,8,10,12\}$,
while Fig.~\ref{M}(b) shows the corresponding extrapolated CCM LSUB$\infty$ ($m \rightarrow
\infty$) results using both Eqs. (\ref{M_extrapo_standard}) and (\ref{M_extrapo_frustrated}).
We note firstly that the CCM LSUB$m$ order parameter results depend on the
approximation level $m$ much more strongly than those for the gs energy.  
It is clear that the order parameter behaves similarly for large values of 
$J_2$ for both the FM model ($J_{1}<0$) and the AFM model ($J_{1}>0$). 
However, once again there are considerable differences in the behavior of $M$ 
between the two models for values of the frustration parameter $J_{2}/|J_{1}| \lesssim 0.6$.  
The extrapolated CCM results for $M$ for the AFM model in Fig.~\ref{M}(b) clearly show the
breakdown of the quasiclassical collinear magnetic LRO near the
critical value of $J_{2} \approx 0.6$, i.e., significantly above the classical
transition point $J^{\rm {cl}}_{2}=0.5$ (and see, e.g.,
Refs.~[\onlinecite{honey2,DJJF:2011_honeycomb,lauchli2011,honey7}]).
Indeed, the CCM estimate for the critical value of the frustration parameter 
in the AFM case for the disappearance of collinear striped order is 
$J^{c_{2}}_{2} \approx 0.60$ from the point at which $M$ becomes zero, using the extrapolation 
scheme of Eq.~(\ref{M_extrapo_frustrated}).\cite{DJJF:2011_honeycomb}
By contrast, the order parameter for the FM model stays almost constant over the whole
parameter region shown in Fig.~\ref{M}.  We do not
observe any indication of the breakdown of the collinear striped
magnetic LRO until $J_{2} \approx 0.11$ for the FM model,
which is below the hypothetical classical transition point $J^{c}_{2} \approx 0.143$, as we 
observed previously in the results for the gs energy.

\begin{figure*}[!tb]
\begin{center}
\mbox{
  \subfloat[CCM LSUB10]{\scalebox{0.3}{\includegraphics[angle=270]{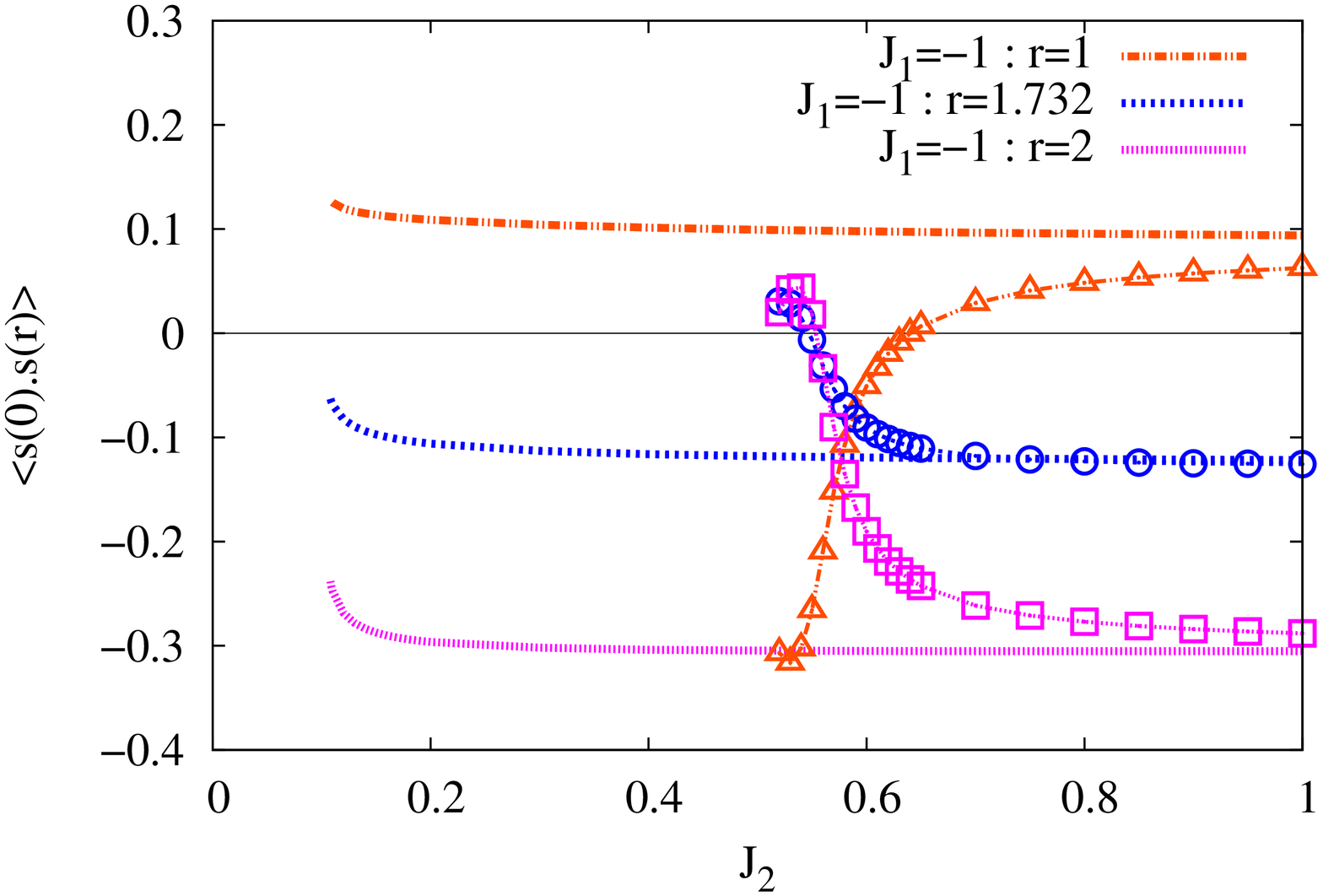}}}
\hfil
\subfloat[ED ($N=32$)]{\scalebox{0.3}{\includegraphics[angle=270]{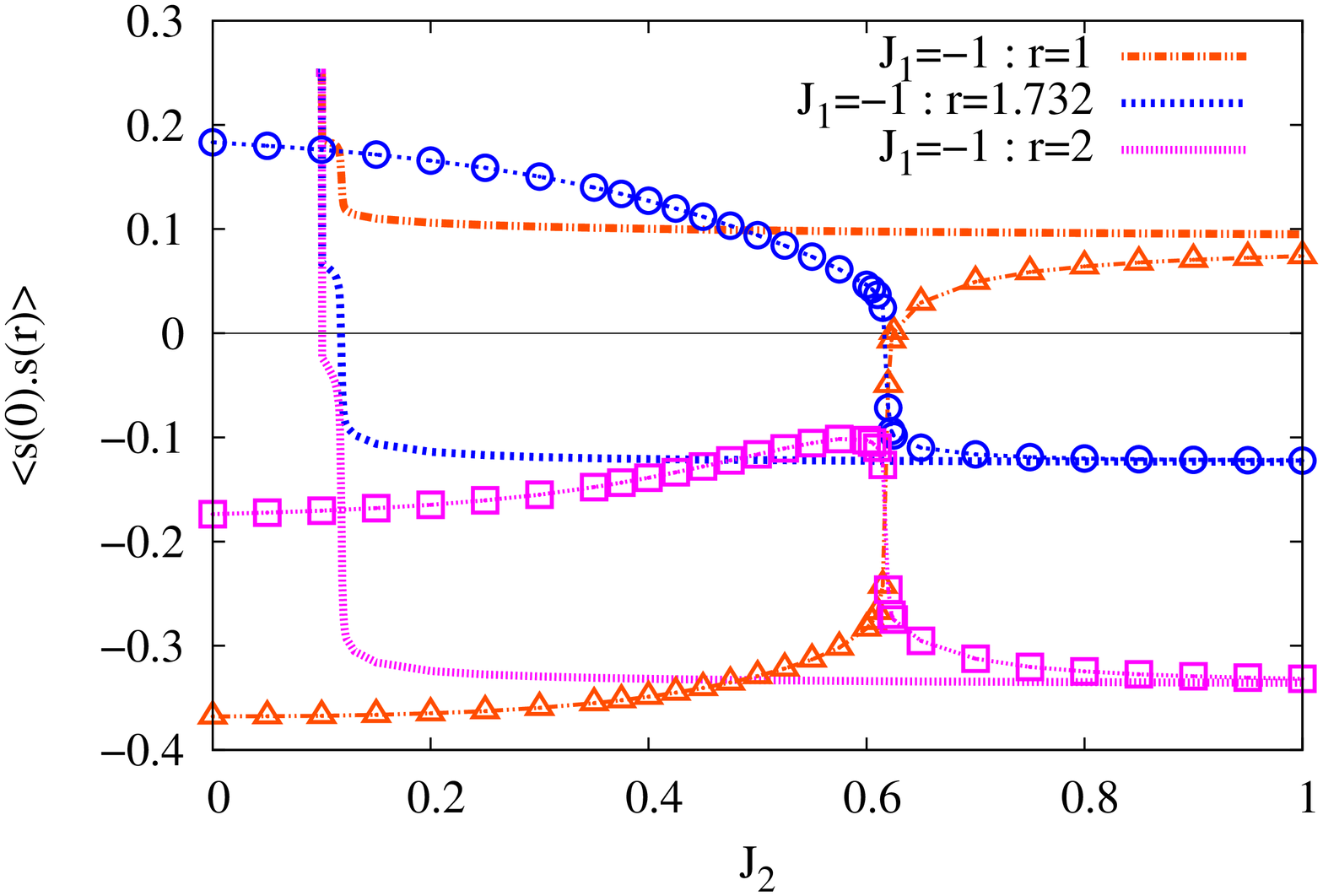}}}
 }
 \caption{
  (Color online) Selected spin-spin correlation functions
  $\langle{\bf s}(0)\cdot{\bf s}({\bf r})\rangle$ for the spin-$\frac{1}{2}$ 
  $J_{1}$--$J_{2}$--$J_{3}$ model on the honeycomb lattice with $J_{3}=J_{2}$, for
  both the FM case (with $J_{1}=-1$) and the AFM case (with $J_{1}=+1$), using
  (a) the CCM with the striped collinear state as model state, at the LSUB$10$ 
  level of approximation; and (b) the ED method on a lattice of size $N=32$.
  Values $r=1, 1.732$ and $2$
  correspond respectively to NN, NNN, and NNNN pairs of spins.  
  In all cases curves without symbols attached refer to the case $J_{1}=-1$,
  whereas the corresponding curves with symbols refer to the case $J_{1}=+1$.}
\label{CorrFunctn}
\end{center}
\end{figure*}

Lastly, we present  results for various spin-spin correlation
functions for the FM as well as for the AFM model in Fig. \ref{CorrFunctn}.
Figure \ref{CorrFunctn}(a) shows the CCM LSUB10 results and 
Fig.~\ref{CorrFunctn}(b) shows the corresponding ED results. 
Once again we note that for large values of the frustration parameter $J_2$
the corresponding spin-spin correlations functions for both the FM ($J_{1}=-1$)
and AFM ($J_{1}=+1$) models agree remarkably well with one another for
both the CCM and ED calculations.  Furthermore, for the FM model the agreement
of the CCM correlation functions with the ED data is excellent.  For the AFM model
the agreement between the CCM and ED results is again excellent for values of $J_2$ above
the transition point at which the AFM collinear striped order disappears, namely 
$J^{c_{2}}_2 \approx 0.60$, but around and below this value there are noticeable
differences.  In particular, the very steep
change in the correlation functions at $J_{2} \approx 0.62$ present in
the ED ($N=32$) data for the AFM model is not observed
in the CCM LSUB10 data. Instead the CCM data show a smoother change in that region.
However, we have argued\cite{DJJF:2011_honeycomb} that for $J < J^{c_{2}}_{2} \approx 0.60$ 
no striped magnetic LRO order exists.  Indeed we argued that no magnetic LRO order 
exists at all for the AFM model in the regime $J^{c_{1}}_{2} < J < J^{c_{2}}_{2}$, where instead
we have a PVBC state.  Hence, it is not
surprising that the CCM solution in a finite order of LSUB$m$
approximation based on the collinear stripe reference state
does not provide such accurate results for the correlation
functions inside this magnetically disordered phase. 

\begin{figure}[!tb]
\begin{center}
\includegraphics[width=6cm,angle=270]{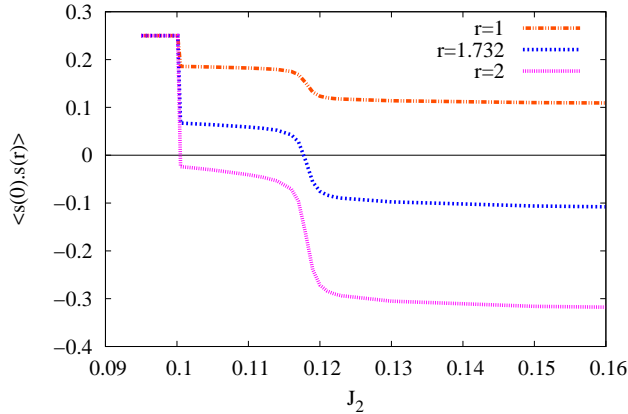}
\caption{(Color online) An expanded view near the FM boundary of the same spin-spin correlation functions 
$\langle{\bf s}(0)\cdot{\bf s}({\bf r})\rangle$ shown in Fig.~\ref{CorrFunctn}(b) 
for the spin-$\frac{1}{2}$ $J_{1}$--$J_{2}$--$J_{3}$ model on the honeycomb lattice for 
the FM case with $J_{1}=-1$ and $J_{3}=J_{2}$ using the ED method on a lattice of size $N=32$.}
\label{CorrFunctn_zoom}
\end{center}
\end{figure}

To conclude, we return to examine more closely the very narrow region 
$0.10 \lesssim J_{2} \lesssim 0.12$ for which our CCM results based on the
striped AFM state as model state could not exclude the possibility of an intervening
phase between the striped AFM and the FM phases.  In Fig.~\ref{CorrFunctn_zoom} 
we show a more detailed view of the ED results for the same spin-spin correlation
functions shown in Fig.~\ref{CorrFunctn}(b) in this narrow region
just above the FM transition point.  The ED data does definitely indicate the
existence of a phase in precisely the region $0.10 \lesssim J_{2} \lesssim 0.12$.
It is difficult from this data to say with any certainty whether or not the state 
is the quantum-mechanical remnant of the classical spiral phase V that exists in the classical 
regime $0.1<J_{2}<0.2$.  Furthermore, without ED calculations on larger lattices, for which the 
computational cost would be prohibitive, it is also not possible to say whether these results
over such a narrow region are an artefact of the finite lattice size.  Our results are summarized in
Sec.~\ref{conclusions}.

\section{Summary and Conclusions}
\label{conclusions}
\begin{figure}[!tb]
\begin{center}
\includegraphics[width=11cm]{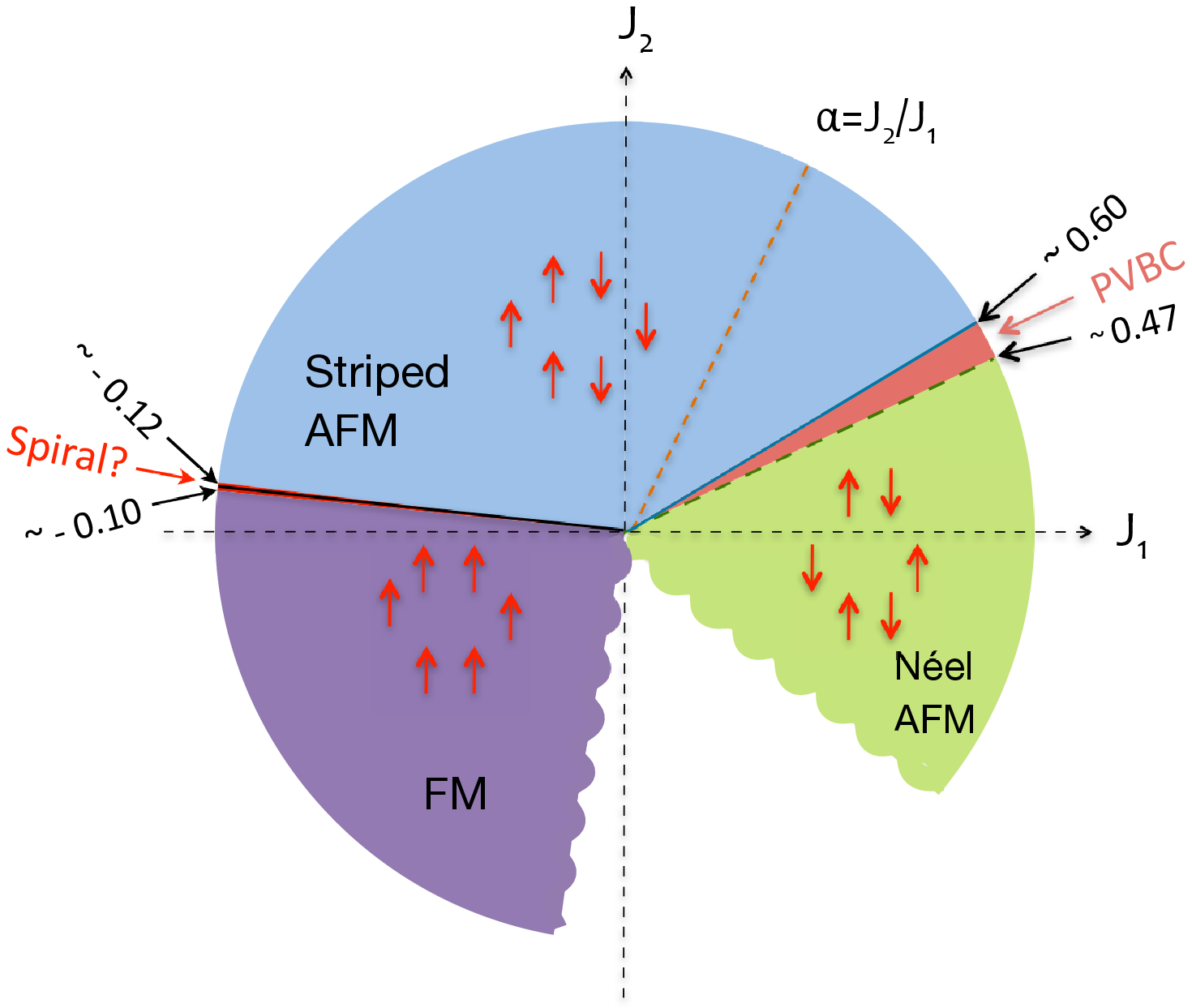}
\caption{(Color online) The phase diagram of the spin-$\frac{1}{2}$ $J_{1}$--$J_{2}$--$J_{3}$ honeycomb model 
in the $J_1$-$J_2$ plane, for the case $J_{3}=J_{2}$.  The continuous transition between the AFM N\'eel  and PVBC phases at
$J_{2}/J_{1} \approx 0.47$ is shown by a broken line, while the first-order transition between the PVBC and AFM striped phases
at $J_{2}/J_{1} \approx 0.60$ is shown by a solid line.  Our results indicate that the transition
between the striped AFM and FM phases is either a first-order one at $J_{2}/J_{1} \approx -0.11$ or occurs via an intermediate
phase, probably with noncollinear spiral order, which exists in the region $-0.12 \lesssim J_{2}/J_{1} \lesssim 0.10$.  
The region between the FM and AFM N\'eel phases with $J_{3}=J_{2}<0$ has not been investigated by us.}
\label{phase-diagram}
\end{center}
\end{figure}

In this paper we have presented results on the gs properties of the
spin-$\frac{1}{2}$ $J_{1}$--$J_{2}$--$J_{3}$ Heisenberg model with FM
NN ($J_{1}=-1$) exchange bonds in the presence of frustrating AFM NNN ($J_{2}>0$) 
and NNNN ($J_{3}>0$) exchange bonds of equal strength ($J_{3}=J_{2}$)
on the honeycomb lattice, using both the CCM and Lanczos ED.
By comparison with previous studies for the AFM ($J_{1}=+1$) 
version of the model,\cite{DJJF:2011_honeycomb} we find similar behavior for 
both models for values $J_{2}\gtrsim 0.6|J_{1}|$, but for values of
$J_{2} \lesssim 0.6|J_{1}|$ the models differ markedly.  The results of the present
paper for the FM version of the model and that of the previous paper \cite{DJJF:2011_honeycomb} 
for the AFM version may conveniently be combined and summarised in the phase diagram
shown in Fig.~\ref{phase-diagram}.

We note that, by contrast with the corresponding model with AFM NN exchange ($J_{1}=+1$)
we do not find indications for a non-classical magnetically-disordered phase
for the model with FM NN exchange ($J_{1}=-1$).  If such a phase exists at all
it must be confined to a very small range of 
the frustration parameter around $0.10 \lesssim J_{2} \lesssim 0.12$.  
However, any such phase is much more likely to be a quasiclassical remnant of the 
spiral phase V that exists in the corresponding classical model (with $J_{1}=-1$) 
in the parameter regime $0.1<J_{2}<0.2$.  As expected, quantum fluctuations then usually favor
a collinear phase over a noncollinear one, and the extent of any spiral phase is
smaller in the quantum spin-$\frac{1}{2}$ case than in the classical ($s \rightarrow 0$)
case.

In one scenario the results presented here for the case $J_{1}=-1$ indicate a direct
first-order transition between the two magnetically ordered
phases, namely  the FM ground state at small values of the frustration parameter $J_2$ 
and  the striped collinear AFM ground state at larger values of $J_2$. 
Our best estimate of the phase transition point is then $J^{c}_{2}=0.1095 \pm 0.0005$.
Although in this scenario a quasiclassical gs phase 
(viz., the collinear striped AFM state) exists in the
whole parameter space down to the FM gs phase, the
frustration might still have a strong effect on 
the low-temperature thermodynamics near the transition
point at $J^{c}_{2}=0.1095 \pm 0.0005$.\cite{haertel10,Richter:2010} 
For values $J_{2} \gtrsim J^{c}_{2}$ the FM multiplet becomes a low-lying
excitation, and this might lead to an additional low-temperature peak
in the specific heat $C(T)$.\cite{shannon04,HMeisner:2006,HTLu:2006}
We note that indications for such  an additional low-temperature peak
in $C(T)$ were also found on the FM side near such a 
transition\cite{haertel08,haertel10} (i.e., at $J_{2}
\lesssim J^{c}_{2}$) in other frustrated spin models.

In an alternative scenario our results also indicate the possibility
of an intervening phase between the collinear FM and striped AFM phases.  Any such phase, 
however, is limited to lie within the very narrow range $0.10 \lesssim J_{2} \lesssim 0.12$,
as shown in Fig.~\ref{phase-diagram}.  In principle we could more accurately establish
the existence of such a phase as a quasiclassical remnant of the classical spiral phase V,
and thence also more accurately establish its phase boundaries, by performing another
comparable set of CCM calculations to those performed here with the striped AFM state as 
model state, but using instead the spiral state V as model state.  Such calculations
would be much more onerous and computationally expensive, however, since on the one hand the number 
$N_f$ of fundamental CCM configurations at a given LSUB$m$ level is greater for the spiral
model state than for the striped model state and, furthermore, the CCM results would need to be
optimized at a given LSUB$m$ level with respect to the spiral pitch angle parameters by minimizing
the corresponding result for the energy per spin separately for each set of values
for the bond strength parameters. 

We note finally that we have not yet investigated the present model in the case where
$J_{3}=J_{2}<0$.  For the FM version of the model when $J_{1}<0$ also, the FM phase
is then obviously the stable ground state.  Conversely, when $J_{1}>0$ and frustration occurs,
there is a direct first-order transition in the classical version of the model between
the FM and N\'eel AFM states at a value $J_{2}/J_{1}=-1$.  Following the discussion in 
Sec.~\ref{results} we might expect that quantum fluctuations could again act either (a) to retain
the direct transition but to stabilize the collinear AFM order in preference to the FM order, 
thus pushing the phase boundary to a somewhat lower value, $J_{2}/J_{1}<-1$, for the spin-$\frac{1}{2}$ 
case; or (b) to permit an intervening state with no classical counterpart.  
Indeed, very preliminary CCM calculations indicate that scenario (a) is realized and
that this corresponding critical point may be pushed to a value $J_{2}/J_{1} \approx -1.15 \pm 0.05$.  
We hope to report in more detail on this region and to give
a more accurate value of this phase boundary in a future paper.

As discussed briefly in Sec.~\ref{intro}, it has been proposed 
\cite{shannon06,shannon09,sindz09,momoi2011} that the competition between FM Heisenberg
interactions between NN pairs of spins and AFM interactions between other spins in
frustrated spin-$\frac{1}{2}$ systems on the square lattice could lead to gapless
spin-liquid states with multipolar order (e.g., spin-nematic states) adjacent 
to the FM state.  Similar states have also been proposed to arise in frustrated multiple 
cyclic spin-exchange models on the triangular lattice with FM NN 
pairwise interactions, \cite{momoi2006} either in the presence of a magnetic field 
(where octupolar order occurs) or in its absence (where quadratic or nematic ordering 
occurs in a state bordering the FM state).  In the case of the frustrated honeycomb-lattice 
ferromagnet considered here we have found no evidence for such states.  However,
the multipolar-ordering phenomenon in the zero-field case considered here is 
evidently rather fragile, and in the square-lattice case 
for the spin-$\frac{1}{2}$ FM version of the $J_1$--$J_2$ model 
(i.e., with $J_{1}<0$) even their existence has 
been questioned in recent rather accurate work \cite{Richter:2010} that also
employed both high-order CCM and ED techniques.  No evidence was found for such states
either in a very recent Schwinger boson study on the square lattice,\cite{Feldner:2011_J1J2J3mod} 
using the same FM version of the spin-$\frac{1}{2}$ $J_{1}$--$J_{2}$--$J_{3}$ 
Heisenberg model that we studied here on ther honeycomb lattice.  
Nevertheless, the history of the study of quantum magnets has shown us 
that the detection of phases 
with novel quantum ordering, such as nematic states of various kinds, is 
extremely subtle.  In particular, the present honeycomb-lattice model surely
warrants further investigation before the absence of nematic states in the 
FM case discussed here is considered definite.

Finally we mention that frustrated ferromagnets are also interesting
with respect to 
multi-magnon bound states appearing in high magnetic fields (and see, e.g., 
Refs.~[\onlinecite{shannon06,Kecke07,Sudan09,nishimoto2011}]).  The
present model also warrants further investigation when the coupling
to an external magnetic field is included.

\section*{ACKNOWLEDGMENTS}
We thank the University of Minnesota Supercomputing Institute for
Digital Simulation and Advanced Computation for the grant of
supercomputing facilities, on which we relied for the
numerical calculations reported here. One of the authors (DJJF) acknowledges and 
thanks the European Science Foundation for financial support under the 
research network program ``Highly Frustrated Magnetism'' (short visit grant number 3858).

\end{document}